\documentstyle[12pt,epsf]{article}
\newcommand{\n}{\hspace*{-2.5mm}}
\newcommand{\gsim}{\;\rlap{\lower 3.5 pt \hbox{$\mathchar \sim$}} \raise 1pt
 \hbox {$>$}\;}
\newcommand{\lsim}{\;\rlap{\lower 3.5 pt \hbox{$\mathchar \sim$}} \raise 1pt
 \hbox {$<$}\;}
\newcommand{\cl}{\mathop{{\mbox{Cl}}_2}\nolimits}
\newcommand{\li}{\mathop{{\mbox{Li}}_4}\nolimits}

\begin{document}
\title{\vskip-3cm{\baselineskip14pt
\centerline{\normalsize\hfill FERMILAB--PUB--95/195--T}
\centerline{\normalsize\hfill MAD/PH/894}
\centerline{\normalsize\hfill MPI/PhT/95--64}
\centerline{\normalsize\hfill TTP95--27\footnote{The complete paper,
including figures, is also available via anonymous ftp at
ftp://ttpux2.physik.uni-karlsruhe.de/, or via www at
http://ttpux2.physik.uni-karlsruhe.de/cgi-bin/preprints/.}}
\centerline{\normalsize\hfill hep-ph/9508241}
\centerline{\normalsize\hfill July 1995}
}
\vskip1.5cm
Three-Loop ${\cal O}(\alpha_s^2G_FM_t^2)$ Corrections to Higgs Production
and Decay at $e^+e^-$ Colliders}
\author{{\sc Bernd A. Kniehl}\thanks{Permanent address:
Max-Planck-Institut f\"ur Physik, Werner-Heisenberg-Institut,
F\"ohringer Ring 6, 80805 Munich, Germany.}\\
{\normalsize Theoretical Physics Department, Fermi National Accelerator
Laboratory,}\\
{\normalsize P.O. Box 500, Batavia, IL 60510, USA}\\
{\normalsize and}\\
{\normalsize Department of Physics, University of Wisconsin,}\\
{\normalsize 1150 University Avenue, Madison, WI~53706, USA}\\ \\
{\sc Matthias Steinhauser}\\
{\normalsize Institut f\"ur Theoretische Teilchenphysik, Universit\"at
Karlsruhe,}\\
{\normalsize Kaiserstra\ss e 12, 76128 Karlsruhe, Germany}}
\date{}
\maketitle
\begin{abstract}
We evaluate the next-to-leading-order QCD corrections of
${\cal O}(\alpha_s^2G_FM_t^2)$ to the Standard-Model $\ell^+\ell^-H$, $ZZH$,
and $W^+W^-H$ couplings in the heavy-top-quark limit.
Exploiting these results together with knowledge of $\Delta\rho$ to the same
order, we analyze a variety of production and decay processes of low-mass Higgs
bosons at $e^+e^-$ colliders.
Specifically, we consider $H\to\ell^+\ell^-$,
$H\to\ell^+\ell^-\ell^{\prime+}\ell^{\prime-}$, $e^+e^-\to ZH$,
$Z\to f\bar fH$, and  $e^+e^-\to f\bar fH$, with $f=\ell,\nu$.
We work in the electroweak on-shell scheme formulated with $G_F$ and employ
both the on-shell and $\overline{\mbox{MS}}$ definitions of the top-quark
mass in QCD.
As expected, the scheme and scale dependences are greatly reduced
when the next-to-leading-order corrections are taken into account.
In the on-shell scheme of top-quark mass renormalization, the
${\cal O}(\alpha_s^2G_FM_t^2)$ corrections act in the same direction as
the ${\cal O}(\alpha_sG_FM_t^2)$ ones and further increase the screening of
the ${\cal O}(G_FM_t^2)$ terms.
The coefficients of $(\alpha_s/\pi)^2$ range from $-6.847$ for the
$ZZH$ coupling to $-16.201$ for the $\ell^+\ell^-H$ coupling.
This is in line with the value $-14.594$ recently found for $\Delta\rho$.
\end{abstract}

\section{Introduction}

After the celebrated discovery of the top quark \cite{abe}, the Higgs boson is
the last missing link in the Standard Model (SM).
The detection of this particle and the study of its characteristics are among
the prime objectives of present and future high-energy colliding-beam
experiments.
Following Bjorken's proposal \cite{bjo}, the Higgs boson is currently being
searched for with the CERN Large Electron-Positron Collider (LEP1) and the
SLAC Linear Collider (SLC) via $e^+e^-\to Z\to f\bar fH$.
At the present time, the failure of this search allows one to rule out the
mass range $M_H\le64.3$~GeV at the 95\% confidence level \cite{jan}.
The quest for the Higgs boson will be continued with LEP2 by exploiting the
Higgs-strahlung mechanism \cite{ell,iof}, $e^+e^-\to ZH\to f\bar fH$.
In next-generation $e^+e^-$ linear supercolliders (NLC), also
$e^+e^-\to\bar\nu_e\nu_eH$ via $W^+W^-$ fusion and, to a lesser extent,
$e^+e^-\to e^+e^-H$ via $ZZ$ fusion will provide copious sources of
Higgs bosons.

Once a novel scalar particle is discovered, it will be crucial to decide if it
is the very Higgs boson of the SM or if it lives in some more extended Higgs
sector.
To that end, precise knowledge of the SM predictions will be mandatory,
{\it i.e.}, quantum corrections must be taken into account.
The status of the radiative corrections to the production and decay processes
of the SM Higgs boson has recently been summarized \cite{pr}.
Since the top quark is by far the heaviest established elementary particle,
with a pole mass of $M_t=(180\pm12)$~GeV \cite{abe}, the leading high-$M_t$
terms, of ${\cal O}(G_FM_t^2)$, are particularly important, and it is desirable
to acquire information on their quantumchromodynamical (QCD) corrections.
During the last year, a number of papers have appeared in which the two-loop
${\cal O}(\alpha_sG_FM_t^2)$ corrections to various Higgs-boson production and
decay processes are presented.
These processes include
$H\to f\bar f$, with $f\ne b$ \cite{hll} and $f=b$ \cite{ks1,kwi},
$Z\to f\bar fH$ and $e^+e^-\to ZH$ \cite{ks2},
$e^+e^-\to\bar\nu_e\nu_eH$ via $W^+W^-$ fusion \cite{ks3},
$gg\to H$ \cite{ks3,gam},
and more \cite{ks3}.
In this paper, we shall take the next step and tackle with three-loop
${\cal O}(\alpha_s^2G_FM_t^2)$ corrections.
To keep matters as simple as possible, we shall restrict our considerations to
light Higgs bosons, with $M_H\ll M_t$, and to reactions with colourless
particles in the initial and final states.
Such reactions typically involve the $\ell^+\ell^-H$, $W^+W^-H$, and $ZZH$
couplings together with gauge couplings of the $W$ and $Z$ bosons to leptons.
We are thus led to incorporate the next-to-leading QCD corrections in the
low-$M_H$ effective $\ell^+\ell^-H$, $W^+W^-H$, and $ZZH$ interaction
Lagrangians.
This will be achieved in Section~2.

Recently, the ${\cal O}(\alpha_s^2G_FM_t^2)$ correction to $\Delta\rho$ has
been calculated and found to be sizeable \cite{avd}.
This is relevant for present and future precision tests of the standard
electroweak theory.
It is of great theoretical interest to find out whether the occurrence of
significant ${\cal O}(\alpha_s^2G_FM_t^2)$ corrections is specific to
$\Delta\rho$ or whether this is a common feature in the class of electroweak
observables with a quadratic $M_t$ dependence at one loop.
In the latter case, there must be some underlying principle which is able to
explain this phenomenon.
Our analysis will put us into a position where we can investigate this matter
for four independent quantities.
We shall return to this issue in Section~5.

The complete evaluation of the one-loop electroweak correction to a process
which involves more than four external particles is enormously intricate.
To our knowledge, the literature does not contain a single example of such a
calculation.
However, the so-called improved Born approximation (IBA) \cite{iba} allows us
to conveniently extract at least the dominant fermionic loop corrections.
As a by-product of our analysis, we shall illustrate the usefulness of the IBA
for Higgs-boson production and decay in high-energy $e^+e^-$ collisions.
The appropriate formalism will be developed in Section~3.

This paper is organized as follows.
In Section~2, we shall extend the low-$M_H$ effective $\ell^+\ell^-H$,
$W^+W^-H$, and $ZZH$ interaction Lagrangians to ${\cal O}(\alpha_s^2G_FM_t^2)$.
In the $G_F$ formulation of the electroweak on-shell scheme, knowledge of the
QCD-corrected $W^+W^-H$ coupling is sufficient to control the related four-
and five-point Higgs-boson production and decay processes which emerge by
connecting one or both of the $W$ bosons with lepton lines, respectively.
Contrariwise, the corresponding processes involving a $ZZH$ coupling receive
additional QCD corrections from the gauge sector, which we shall evaluate by
invoking the IBA in Section~3.
In Section~4, we shall quantitatively analyze the phenomenological consequences
of our results.
Section~5 contains our conclusions.

\section{Effective Lagrangians}

Throughout this paper, we shall work in the electroweak on-shell
renormalization scheme \cite{aok}, with $G_F$ as a basic parameter, and define
$c_w^2=1-s_w^2=M_W^2/M_Z^2$ \cite{sir}.
In particular, this implies that the lowest-order formulae are expressed in
terms of $G_F$, $c_w$, $s_w$, and the physical particle masses.
The self-energies of the $W$, $Z$, and Higgs bosons to
${\cal O}(\alpha_s^2G_FM_t^2)$ for zero external four-momentum squared
will be the basic ingredients of our analysis.
While the results for the $W$ and $Z$ bosons are now well established
\cite{avd}, the Higgs-boson self-energy requires a separate analysis,
which will be performed here.
Our calculation will proceed along the lines of Ref.~\cite{avd}.
We shall employ dimensional regularization in $n=4-2\epsilon$ space-time
dimensions and introduce a 't~Hooft mass, $\mu$, to keep the coupling constants
dimensionless.
We shall suppress terms containing $\gamma_E-\ln(4\pi)$, where $\gamma_E$ is
Euler's constant.
These terms may be retrieved by substituting
$\mu^2\to4\pi e^{-\gamma_E}\mu^2$.
In the modified minimal-subtraction ($\overline{\mbox{MS}}$) scheme \cite{msb},
these terms are subtracted along with the poles in $\epsilon$.
This is also true for the relation between the $\overline{\mbox{MS}}$ and pole
masses of the quarks, so that these terms are also absent when the quark masses
are renormalized according to the on-shell scheme.
Since we wish to extract the leading high-$M_t$ terms, we may neglect the
masses of all virtual particles, except for the top quark.
As usual, we shall take $\gamma_5$ to be anticommuting for $n$ arbitrary.
We shall choose a covariant gauge with an arbitrary gauge parameter
for the gluon propagator.
This will allow us to explicitly check that our final results are gauge
independent.
The requirement that the expressions for physical observables be
renormalization-group (RG) invariant will serve as a further check for our
calculation.

Large intermediate expressions will be treated with the help of FORM 2.0
\cite{ver}.
The tadpole integrals which enter the one- and two-loop calculations may be
solved straightforwardly, even for arbitrary powers of propagators.
The three-loop case is more involved.
After evaluating the traces, the scalar integrals may be reduced by
decomposing the scalar products in the numerator into appropriate combinations
of the factors in the denominator.
Subsequently, recurrence relations derived using the integration-by-parts
method \cite{che} may be applied to reduce any scalar Feynman integral to a
small number of so-called master diagrams, which remain to be calculated by
hand.
More technical details may be found in Ref.~\cite{avd}.

Prior to listing our results, we shall introduce our notation.
We take the colour gauge group to be SU($N_c$);
$C_F=(N_c^2-1)/(2N_c)$ and $C_A=N_c$ are the Casimir operators of its
fundamental and adjoint representations, respectively.
As is usually done for SU($N_c$), we fix the trace normalization of the
fundamental representation to be $T_F=1/2$.
In our numerical analysis, we set $N_c=3$.
We explicitly include five massless quark flavours plus the massive top quark
in our calculation, so that we have $n_f=6$ active quark flavours altogether,
{\it i.e.}, we must not consider $n_f$ as a free parameter.
We denote the QCD renormalization scale by $\mu$.
We evaluate the strong coupling constant, $\alpha_s(\mu)$, at next-to-leading
order (two loops) in the $\overline{\mbox{MS}}$ scheme, from
\begin{equation}
\label{as}
{\alpha_s(\mu)\over\pi}={1\over\beta_0\ln(\mu^2/\Lambda_{\overline{MS}}^2)}
\left[1-{\beta_1\over\beta_0^2}\,{\ln\ln(\mu^2/\Lambda_{\overline{MS}}^2)
\over\ln(\mu^2/\Lambda_{\overline{MS}}^2)}\right],
\end{equation}
where $\Lambda_{\overline{MS}}$ is the asymptotic scale parameter appropriate
for $n_f=6$ and \cite{wil}
\begin{eqnarray}
\label{beta}
\beta_0&\n=\n&{1\over4}\left({11\over3}C_A-{2\over3}n_f\right)={7\over4},
\nonumber\\
\beta_1&\n=\n&{1\over16}\left({34\over3}C_A^2-2C_Fn_f-{10\over3}C_An_f\right)
={13\over8}
\end{eqnarray}
are the first two coefficients of the Callan-Symanzik beta function of QCD.
We define $a=4h=\alpha_s(\mu)/\pi$,
$x_t=[G_Fm_t^2(\mu)/8\pi^2\sqrt2]$,
$X_t=(G_FM_t^2/8\pi^2\sqrt2)$,
$l=\ln[\mu^2/m_t^2(\mu)]$, and
$L=\ln(\mu^2/M_t^2)$,
where $m_t(\mu)$ and $M_t$ are the $\overline{\mbox{MS}}$ and pole masses of
the top quark, respectively, and $G_F$ is Fermi's constant.
Using the two-loop relation between $m_t(M_t)$ and $M_t$ \cite{gra} along with
the RG equation for $m_t(\mu)$, we find
\begin{eqnarray}
\label{mass}
{m_t(\mu)\over M_t}&\n=\n&1+hC_F(-3L-4)
+h^2C_F\left\{L^2\left({9\over2}C_F-{11\over2}C_A+n_f\right)
+L\left({21\over2}C_F-{185\over6}C_A
\right.\right.\nonumber\\
&\n+\n&\left.
{13\over3}n_f\right)-12\zeta(2)+6
+C_F\left[-12\zeta(3)+6\zeta(2)(8\ln2-5)+{7\over8}\right]
\nonumber\\
&\n+\n&\left.
C_A\left[6\zeta(3)+8\zeta(2)(-3\ln2+1)-{1111\over24}\right]
+n_f\left[4\zeta(2)+{71\over12}\right]\right\}
\nonumber\\
&\n\approx\n&1-a\left(L+{4\over3}\right)
-a^2\left({3\over8}L^2+{35\over8}L+9.125\,451\right).
\end{eqnarray}
Riemann's zeta function takes on the values $\zeta(2)=\pi^2/6$,
$\zeta(3)\approx1.202\,057$, and $\zeta(4)=\pi^4/90$.
The numerical constants \cite{avd}
\begin{eqnarray}
S_2&\n=\n&{4\over9\sqrt3}\cl\left({\pi\over3}\right)
\approx0.260\,434,\nonumber\\
D_3&\n\approx\n&-3.027\,009,\nonumber\\
B_4&\n=\n&16\li\left({1\over2}\right)-{13\over2}\zeta(4)-4\zeta(2)\ln^22
+{2\over3}\ln^42
\approx-1.762\,800,
\end{eqnarray}
where $\cl{}$ is Clausen's function and $\li{}$ is the quadrilogarithm,
occur in the evaluation of the three-loop master diagrams.

In the following , we shall frequently make use of the QCD expansion of
$\Delta\rho$ through ${\cal O}(\alpha_s^2G_FM_t^2)$.
{}For the reader's convenience, we shall list it here for $N_c=3$ and $n_f=6$.
The $\overline{\mbox{MS}}$ and on-shell results read \cite{avd}
\begin{eqnarray}
\Delta\bar\rho&\n\approx\n&
N_cx_t\left[1+a(2\,l-0.193\,245)
+a^2\left({15\over4}l^2+2.025\,330\,l-3.969\,560\right)\right],
\\
\label{drhoos}
\Delta\rho&\n\approx\n&
N_cX_t[1-2.859\,912\,a-a^2(5.004\,846\,L+14.594\,028)],
\end{eqnarray}
respectively.

To start with, we shall construct the low-$M_H$ effective $\ell^+\ell^-H$
interaction Lagrangian through ${\cal O}(\alpha_s^2G_FM_t^2)$.
In the following, bare quantities will be labelled with the superscript 0.
The bare $\ell^+\ell^-H$ Lagrangian reads
\begin{equation}
\label{llh}
{\cal L}_{\ell\ell H}=-m_\ell^0\bar\ell^0\ell^0{H^0\over v^0},
\end{equation}
where $v$ denotes the Higgs vacuum expectation value.
The renormalizations of the lepton mass and wave function do not receive
corrections in ${\cal O}(\alpha_s^nG_FM_t^2)$, where $n=0,1,2$,
so that we may replace $m_\ell^0$ and $\ell^0$ with their renormalized
counterparts.
In the $G_F$ formulation of the on-shell scheme, we have \cite{hff}
\begin{equation}
\label{du}
{H^0\over v^0}=2^{1/4}G_F^{1/2}H(1+\delta_u),
\end{equation}
with
\begin{equation}
\delta_u=-{1\over2}\left[{\Pi_{WW}(0)\over M_W^2}+\Pi_{HH}^\prime(0)\right].
\end{equation}
Here, $\Pi_{WW}(q^2)$ and $\Pi_{HH}(q^2)$ are the $W$- and Higgs-boson
self-energies for external momentum $q$, respectively, and the subscript $u$
is to remind us that this term appears as a universal building block in the
radiative corrections to all production and decay processes of the Higgs boson.
Consequently, the renormalized version of Eq.~(\ref{llh}) reads
\begin{equation}
\label{lllh}
{\cal L}_{\ell\ell H}=-2^{1/4}G_F^{1/2}m_\ell\bar\ell\ell H(1+\delta_u).
\end{equation}

\begin{figure}[ht]
 \begin{center}
 \begin{tabular}{ccc}
   \epsfxsize=5.0cm
   \leavevmode
   \epsffile[130 260 470 530]{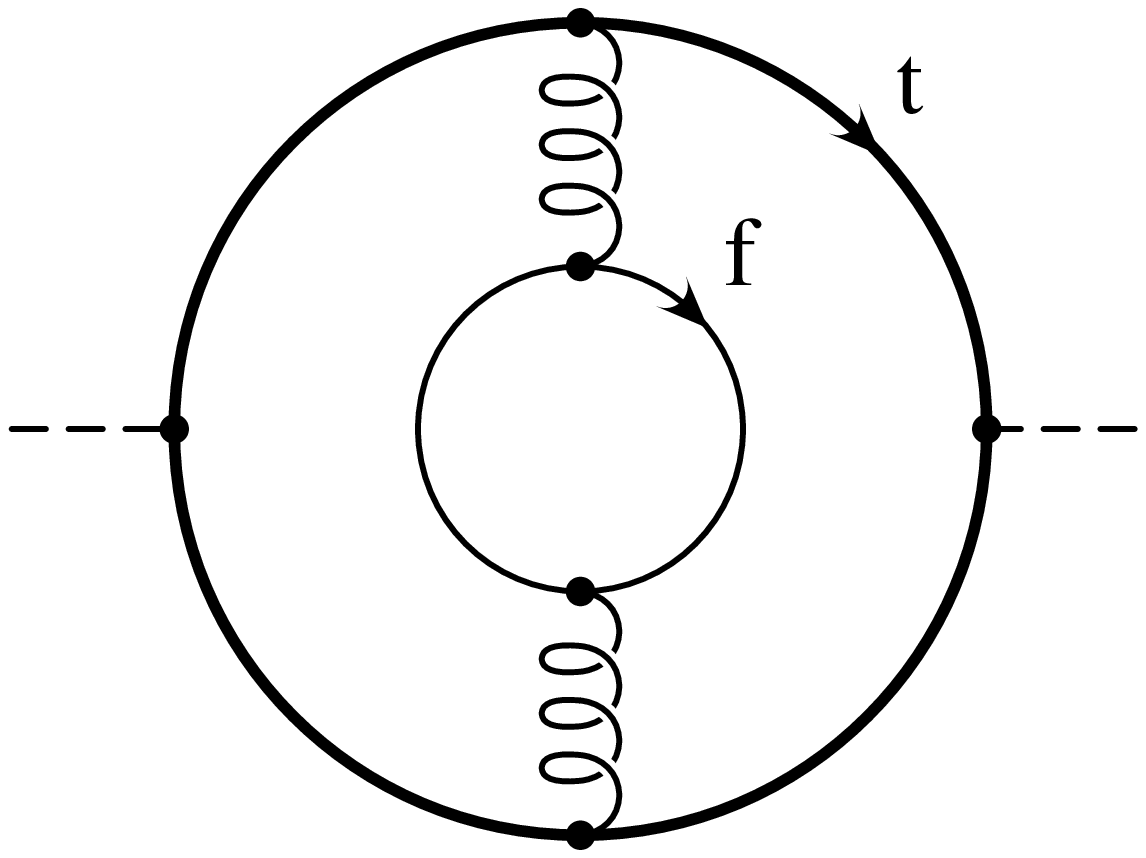}
   &
   \epsfxsize=5.0cm
   \leavevmode
   \epsffile[130 260 470 530]{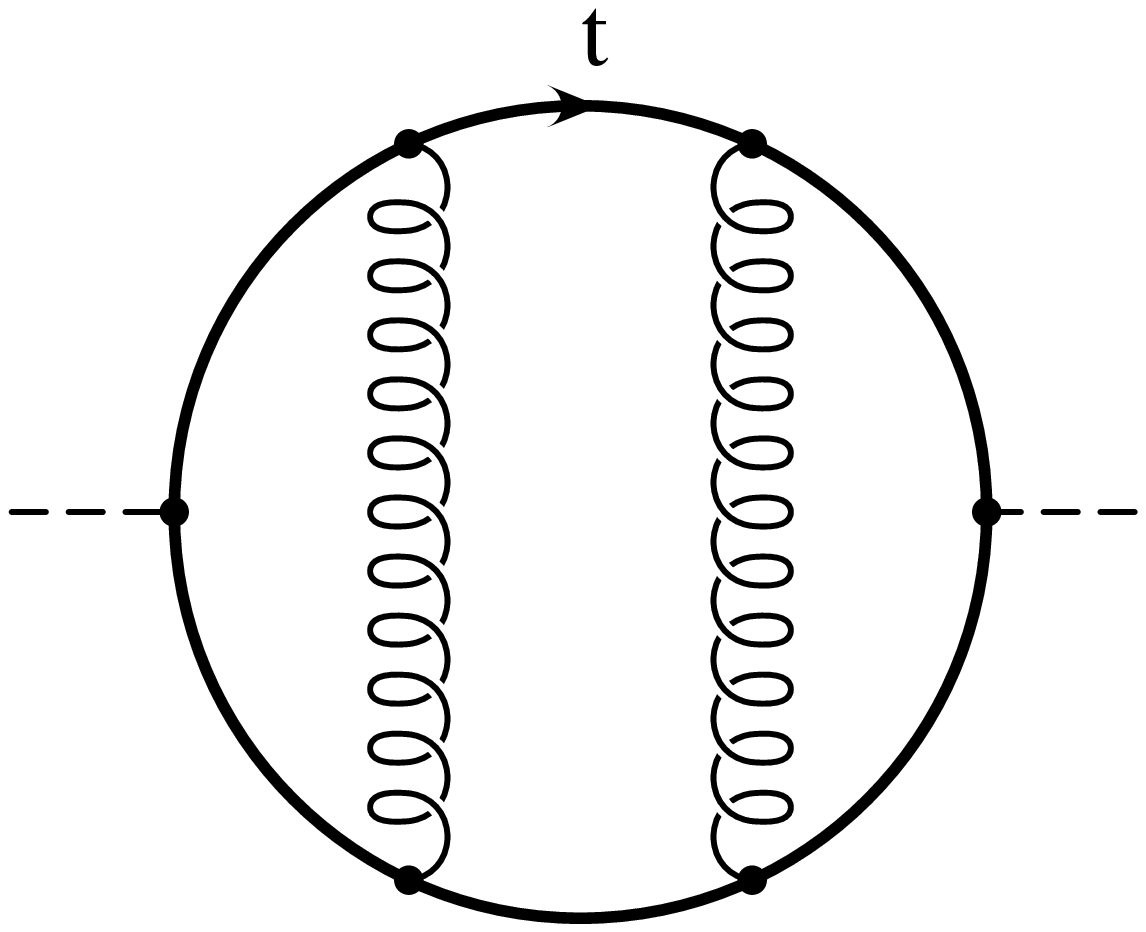}
   &
   \epsfxsize=5.0cm
   \leavevmode
   \epsffile[130 260 470 530]{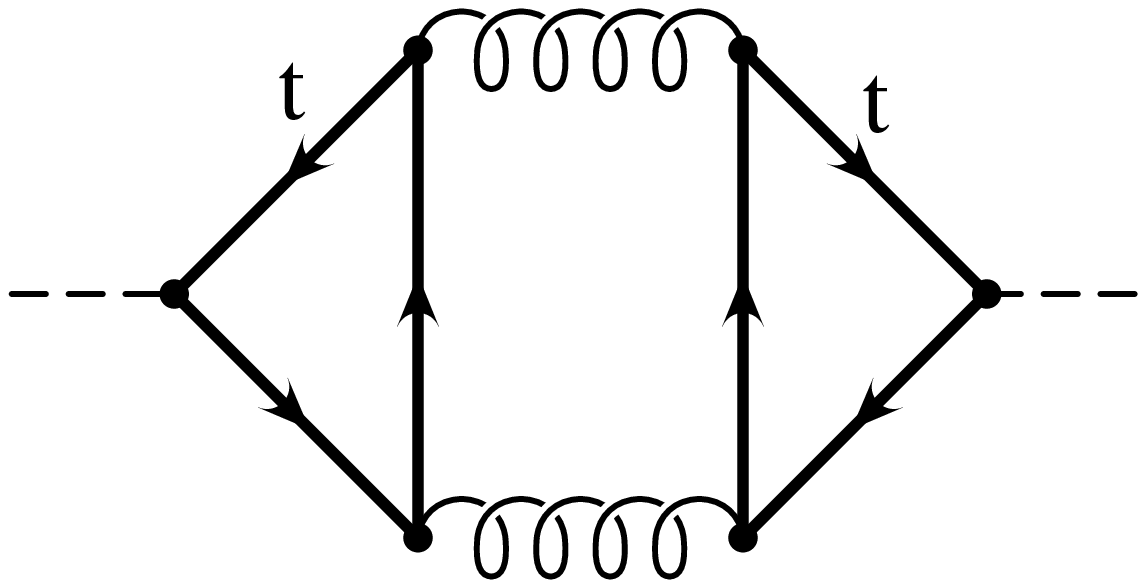}
 \end{tabular}
  \caption{\label{one}Typical Feynman diagrams pertinent to $\Pi_{HH}(q^2)$ in
  ${\cal O}(\alpha_s^2G_FM_t^2)$.
  $f$ stands for any quark.}
 \end{center}
\end{figure}

The one-loop expressions for $\Pi_{WW}(q^2)$ and $\Pi_{HH}(q^2)$ have been
presented in Ref.~\cite{hzz}.
The leading-order QCD corrections to $\Pi_{WW}(q^2)$ and $\Pi_{HH}(q^2)$ for
arbitrary quark masses have been found in Refs.~\cite{djo,hll}, respectively.
The ${\cal O}(\alpha_sG_FM_t^2)$ term of $\delta_u$ has independently been
obtained in Ref.~\cite{kwi} by using the computational technique outlined
above at the two-loop level.
Here, we shall extend this analysis to ${\cal O}(\alpha_s^2G_FM_t^2)$.
The ${\cal O}(\alpha_s^2G_FM_t^2)$ term of $\Pi_{WW}(0)$ may be found in
Ref.~\cite{avd}.
The Feynman diagrams pertinent to $\Pi_{HH}(q^2)$ in
${\cal O}(\alpha_s^2G_FM_t^2)$ come in twenty different topologies.
Typical examples are depicted in Fig.~\ref{one}.
We shall renormalize the strong coupling constant and the top-quark mass
according to the $\overline{\mbox{MS}}$ scheme.
The appropriate counterterms are listed in Ref.~\cite{gra}.
In this way, we obtain
\begin{eqnarray}
\label{pihh}
\Pi_{HH}^\prime(0)&\n=\n&N_cx_t\left\{{2\over\epsilon}+2l-{4\over3}
+hC_F\left(-{6\over\epsilon^2}+{5\over\epsilon}+6l^2-10l-{37\over6}\right)
+h^2C_F\left[27\zeta(3)+6
\right.\right.
\nonumber\\
&\n+\n&C_F\left({12\over\epsilon^3}-{12\over\epsilon^2}
+{1\over\epsilon}\left(24\zeta(3)-{119\over6}\right)
+l\left(72\zeta(3)-{93\over2}\right)+24B_4-108\zeta(4)
\right.\nonumber\\
&\n+\n&\left.
106\zeta(3)+{331\over12}\right)
+C_A\left({22\over3\epsilon^3}-{83\over3\epsilon^2}
+{1\over\epsilon}\left(-12\zeta(3)+{77\over3}\right)
+{22\over3}l^3
\right.\nonumber\\
&\n+\n&\left.
14l^2+l\left(-36\zeta(3)-{961\over18}\right)-12B_4+54\zeta(4)
-{55\over3}\zeta(3)-7\right)
\nonumber\\
&\n+\n&\left.\left.
n_f\left(-{4\over3\epsilon^3}+{10\over3\epsilon^2}-{8\over3\epsilon}
-{4\over3}l^3+{65\over9}l-{32\over3}\zeta(3)-3\right)\right]\right\}.
\end{eqnarray}
When we combine Eq.~(\ref{pihh}) with the corresponding expression for
$\Pi_{WW}(0)$ \cite{avd}, the ultraviolet divergences cancel, and we obtain
\begin{eqnarray}
\bar\delta_u&\n=\n&N_cx_t\left\{{7\over6}
+hC_F\left(7l-2\zeta(2)+{19\over3}\right)
+h^2C_F\left[243S_2-{449\over6}\zeta(3)-{14\over3}\zeta(2)+{79\over3}
\right.\right.
\nonumber\\
&\n+\n&C_F\left(21l^2+l\left(-12\zeta(2)-{1\over2}\right)+4B_4+2D_3
-{1053\over2}S_2+2\zeta(4)+{599\over3}\zeta(3)-{259\over9}\zeta(2)
\right.\nonumber\\
&\n-\n&\left.
{3043\over72}\right)
+C_A\left({77\over6}l^2+l\left(-{22\over3}\zeta(2)+{1097\over18}\right)
-2B_4-D_3+{1053\over4}S_2+15\zeta(4)
\right.\nonumber\\
&\n-\n&\left.\left.\left.\!\!
{509\over6}\zeta(3)
-{73\over3}\zeta(2)
+{953\over24}\right)
+n_f\left(\!-{7\over3}l^2+l\left({4\over3}\zeta(2)-{73\over9}\right)
\!-{8\over3}\zeta(3)+{14\over3}\zeta(2)-{55\over12}\right)\right]\right\}
\nonumber\\
&\n\approx\n&
{7\over6}N_cx_t\left[1+a(2\,l+0.869\,561)
+a^2\left({15\over4}l^2+6.010\,856\,l-2.742\,226\right)\right].
\end{eqnarray}
With the help of Eq.~(\ref{mass}), we may eliminate $m_t(\mu)$ in favour of
$M_t$, which leads to
\begin{eqnarray}
\label{duos}
\delta_u&\n=\n&N_cX_t\left\{{7\over6}
+hC_F\left(-2\zeta(2)-3\right)
+h^2C_F\left[243S_2-{449\over6}\zeta(3)-{98\over3}\zeta(2)+{121\over3}
\right.\right.
\nonumber\\
&\n+\n&C_F\left(4B_4+2D_3-{1053\over2}S_2+2\zeta(4)+{515\over3}\zeta(3)
+\zeta(2)\left(112\ln2-{745\over9}\right)-{146\over9}\right)
\nonumber\\
&\n+\n&C_A\left(L\left(-{22\over3}\zeta(2)-11\right)
-2B_4-D_3+{1053\over4}S_2+15\zeta(4)-{425\over6}\zeta(3)
\right.\nonumber\\
&\n+\n&\left.\left.\left.
\zeta(2)\left(-56\ln2-{17\over3}\right)-{2459\over36}\right)
+n_f\left(L\left({4\over3}\zeta(2)+2\right)
-{8\over3}\zeta(3)+14\zeta(2)+{83\over9}\right)\right]\right\}
\nonumber\\
&\n\approx\n&{7\over6}N_cX_t[1-1.797\,105\,a-a^2(3.144\,934\,L+16.200\,847)].
\end{eqnarray}
Equation~(\ref{duos}) reproduces the ${\cal O}(G_FM_t^2)$ and
${\cal O}(\alpha_sG_FM_t^2)$ terms found in Refs.~\cite{hff,hll}, respectively.
We observe that the new ${\cal O}(\alpha_s^2G_FM_t^2)$ term in Eq.~(\ref{duos})
enhances the QCD correction and thus supports the screening of the
leading-order $M_t$ dependence.
The choice $\mu=M_t$ is singled out, since it eliminates the terms containing
$L$ in Eq.~(\ref{duos}).
The nonlogarithmic coefficient of $(\alpha_s/\pi)^2$ in Eq.~(\ref{duos}) is
relatively large;
it exceeds the corresponding coefficient of $\Delta\rho$ in Eq.~(\ref{drhoos})
by approximately 11\%.
If we consider the ratio of the coefficient of $(\alpha_s/\pi)^2$ to the one of
$\alpha_s/\pi$, the difference is even more pronounced;
the corresponding numbers for Eqs.~(\ref{duos}) and (\ref{drhoos}) are
roughly 9 versus 5.

A phenomenologically interesting application of Eq.~(\ref{lllh}) is to study
the effect of QCD corrections on $\Gamma(H\to\ell^+\ell^-)$.
The corrections through ${\cal O}(\alpha_s^2G_FM_t^2)$ to this observable may
be accommodated by multiplying the Born formula \cite{hff} with
\begin{eqnarray}
\label{kllh}
K_{\ell\ell H}&\n=\n&(1+\delta_u)^2\nonumber\\
&\n=\n&1+2\delta_u,
\end{eqnarray}
where we have suppressed terms of ${\cal O}(G_F^2M_t^4)$ in the second line.
This implies that $\delta_u$ is gauge independent and RG invariant in these
orders.
In order to avoid double counting, the ${\cal O}(G_FM_t^2)$ term must once be
subtracted when the full one-loop correction \cite{hff} is included.
A detailed numerical analysis will be presented in Section~4.

Next, we shall derive the ${\cal O}(\alpha_s^2G_FM_t^2)$ correction to the
low-$M_H$ effective $W^+W^-H$ interaction Lagrangian.
In contrast to the $\ell^+\ell^-H$ case, we are now faced with the task of
computing genuine three-point amplitudes at three loops, which, at first sight,
appears to be enormously hard.
Fortunately, in the limit that we are interested in, this problem may be
reduced to one involving just three-loop two-point diagrams by means of a
low-energy theorem \cite{ell,vai}.
Generally speaking, this theorem relates the amplitudes of two processes which
differ by the insertion of an external Higgs-boson line carrying zero
four-momentum.
It allows us to compute a loop amplitude, ${\cal M}(A\to B+H)$, with an
external Higgs boson which is light compared to the virtual particles by
differentiating the respective amplitude without that Higgs boson,
${\cal M}(A\to B)$, with respect to the virtual-particle masses.
More precisely \cite{ell,vai},
\begin{equation}
\label{let}
\lim_{p_H\to0}{\cal M}(A\to B+H)={1\over v}\sum_i
{m_i\partial\over\partial m_i}{\cal M}(A\to B),
\end{equation}
where $i$ runs over all massive virtual particles which are involved in the
transition $A\to B$.
Here, it is understood that the differential operator does not act on factors
of $m_i$ appearing in coupling constants, since this would generate tree-level
interactions involving the Higgs boson that do not exist in the SM.
This theorem has variously been applied at leading order \cite{ell,vai} and has
even made its way into standard text books \cite{oku}.
Special care must be exercised if this theorem is to be applied beyond leading
order.
Then, it must be formulated for the bare quantities of the theory, and the
renormalization must be performed after the left-hand side of Eq.~(\ref{let})
has been constructed \cite{ks1}.
The beyond-leading-order version of this theorem \cite {ks1} has recently been
employed to find the ${\cal O}(\alpha_sG_FM_t^2)$ corrections to
$\Gamma\left(H\to b\bar b\,\right)$ \cite{ks1},
$\Gamma\left(Z\to f\bar fH\right)$, and $\sigma(e^+e^-\to ZH)$ \cite{ks2}.
A comprehensive review of higher-order applications of this and related
low-energy theorems may be found in Ref.~\cite{ks3}.
An axiomatic formulation of these soft-Higgs theorems has recently been
introduced in Ref.~\cite{kil}.

Proceeding along the lines of Refs.~\cite{ks2,ks3}, we find the bare $W^+W^-H$
interaction Lagrangian including its genuine vertex corrections to be
\begin{equation}
{\cal L}_{W^+W^-H}=2(M_W^0)^2(W_\mu^+)^0(W^{-\mu})^0{H^0\over v^0}
\left[1-{(m_t^0)^2\partial\over\partial(m_t^0)^2}\,
{\Pi_{WW}(0)\over(M_W^0)^2}\right],
\end{equation}
where it is understood that $\Pi_{WW}(0)$ is expressed in terms of the bare
top-quark mass, $m_t^0$, while all other quark masses are put to zero.
We renormalize the $W$-boson mass and wave function
according to the electroweak on-shell scheme by substituting
\begin{eqnarray}
(M_W^0)^2&\n=\n&M_W^2+\delta M_W^2,\nonumber\\
(W_\mu^\pm)^0&\n=\n&W_\mu^\pm(1+\delta Z_W)^{1/2},
\end{eqnarray}
with the counterterms
\begin{eqnarray}
\delta M_W^2&\n=\n&\Pi_{WW}(0),\nonumber\\
\delta Z_W&\n=\n&-\Pi_{WW}^\prime(0).
\end{eqnarray}
{}For dimensional reasons, $\delta Z_W$ does not receive corrections in the
orders that we are interested in.
Using Eq.~(\ref{du}), we thus obtain
\begin{equation}
\label{lwwh}
{\cal L}_{W^+W^-H}=2^{5/4}G_F^{1/2}M_W^2W_\mu^+W^{-\mu}H(1+\delta_{WWH}),
\end{equation}
where
\begin{equation}
\delta_{WWH}=\delta_u+\delta_{nu}^{WWH}
\end{equation}
and the non-universal part herein may be calculated from
\begin{equation}
\label{dwwhnu}
\delta_{nu}^{WWH}=\left[1-{(m_t^0)^2\partial\over\partial(m_t^0)^2}\right]
{\Pi_{WW}(0)\over(M_W^0)^2}.
\end{equation}
In Ref.~\cite{avd}, $\Pi_{WW}(0)$ is expressed in terms of renormalized
parameters.
Thus, we have to undo the top-quark mass renormalization \cite{gra} before we
can apply Eq.~(\ref{dwwhnu}).
Then, after evaluating the right-hand side of Eq.~(\ref{dwwhnu}), we
reintroduce the renormalized top-quark mass and so obtain a finite result for
$\delta_{nu}^{WWH}$, which we combine with $\delta_u$ to get $\delta_{WWH}$.
If we define the top-quark mass according to the $\overline{\mbox{MS}}$
scheme, then the result is
\begin{eqnarray}
\bar\delta_{WWH}&\n=\n&N_cx_t\left\{-{5\over6}
+hC_F\left(-5l-2\zeta(2)+{7\over3}\right)
+h^2C_F\left[243S_2-{449\over6}\zeta(3)-{14\over3}\zeta(2)+{79\over3}
\right.\right.
\nonumber\\
&\n+\n&C_F\left(-15l^2+l\left(-12\zeta(2)+{83\over2}\right)+4B_4+2D_3
-{1053\over2}S_2+2\zeta(4)+{383\over3}\zeta(3)
\right.\nonumber\\
&\n-\n&\left.
{43\over9}\zeta(2)
+{377\over72}\right)
+C_A\left(-{55\over6}l^2+l\left(-{22\over3}\zeta(2)-{331\over18}\right)
-2B_4-D_3+{1053\over4}S_2
\right.\nonumber\\
&\n+\n&\left.
15\zeta(4)-{293\over6}\zeta(3)-{29\over3}\zeta(2)
-{691\over24}\right)
+n_f\left({5\over3}l^2+l\left({4\over3}\zeta(2)+{11\over9}\right)
-{8\over3}\zeta(3)
\right.\nonumber\\
&\n+\n&\left.\left.\left.
2\zeta(2)+{53\over12}\right)\right]\right\}
\nonumber\\
&\n\approx\n&
-{5\over6}N_cx_t\left[1+a(2\,l+0.382\,614)
+a^2\left({15\over4}l^2+4.184\,802\,l+1.343\,710\right)\right].
\end{eqnarray}
The corresponding formula written in terms of $M_t$ reads
\begin{eqnarray}
\label{dwwhos}
\delta_{WWH}&\n=\n&N_cX_t\left\{-{5\over6}
+hC_F\left(-2\zeta(2)+9\right)
+h^2C_F\left[243S_2-{449\over6}\zeta(3)+{46\over3}\zeta(2)+{49\over3}
\right.\right.\nonumber\\
&\n+\n&C_F\left(4B_4+2D_3-{1053\over2}S_2+2\zeta(4)+{443\over3}\zeta(3)
+\zeta(2)\left(-80\ln2+{551\over9}\right)-{614\over9}\right)
\nonumber\\
&\n+\n&C_A\left(L\left(-{22\over3}\zeta(2)+33\right)
-2B_4-D_3+{1053\over4}S_2+15\zeta(4)-{353\over6}\zeta(3)
\right.\nonumber\\
&\n+\n&\left.\left.\left.
\zeta(2)(40\ln2-23)+{1741\over36}\right)
+n_f\left(L\left({4\over3}\zeta(2)-6\right)
-{8\over3}\zeta(3)-{14\over3}\zeta(2)-{49\over9}\right)\right]\right\}
\nonumber\\
&\n\approx\n&-{5\over6}N_cX_t[1-2.284\,053\,a-a^2(3.997\,092\,L+10.816\,384)].
\end{eqnarray}
We recover the ${\cal O}(G_FM_t^2)$ and ${\cal O}(\alpha_sG_FM_t^2)$ terms of
Refs.~\cite{hww,ks3}, respectively.
Similarly to $\Delta\rho$ and $\delta_u$, the ${\cal O}(\alpha_s^2G_FM_t^2)$
term of Eq.~(\ref{dwwhos}) supports the screening of the one-loop $M_t$
dependence by the leading-order QCD correction.
Here, the coefficient of $(\alpha_s/\pi)^2$ is by 26\% smaller than in the
case of $\Delta\rho$, but it, too, is about five times bigger than the
coefficient of $\alpha_s/\pi$.

{}From Eq.~(\ref{lwwh}) it follows on that $\Gamma(H\to W^+W^-)$ receives the
correction factor
\begin{equation}
\label{kwwh}
K_{WWH}=1+2\delta_{WWH}.
\end{equation}
Thus, both $\delta_{nu}^{WWH}$ and $\delta_{WWH}$ are gauge independent and RG
invariant to the orders that we are working in.
The tree-level formula for $\Gamma(H\to W^+W^-)$ and its full one-loop
correction may be found in Ref.~\cite{hww}.
In order for the Higgs boson to decay into a $W^+W^-$ pair, it must satisfy
$M_H>2M_W$.
On the other hand, the high-$M_t$ approximation is based on $M_H\ll M_t$.
Since these two conditions conflict with each other \cite{abe}, the
application of Eq.~(\ref{lwwh}) to $\Gamma(H\to W^+W^-)$ is somewhat academic.
However, the first condition is relaxed to $M_H>M_W$ or removed altogether
if one or both of the $W$ bosons are allowed to leave their mass shells,
respectively.
In order to avoid gluon exchange between the $W^+W^-H$ vertex and the external
fermions, we restrict our considerations to leptonic currents.
The resulting class of processes includes
$H\to(W^+)^*W^-\to\ell^+\nu_\ell W^-$,
$H\to W^+(W^-)^*\to W^+\ell^-\bar\nu_\ell$,
$H\to(W^+)^*(W^-)^*\to\ell^+\nu_\ell\ell^{\prime-}\bar\nu_{\ell^\prime}$,
as well as $e^+e^-\to\bar\nu_e\nu_e(W^+)^*(W^-)^*\to\bar\nu_e\nu_eH$ via
$W^+W^-$ fusion.
The Born formulae for these $1\to3$, $1\to4$, and $2\to3$ processes may be
found in Refs.~\cite{riz,hzgg,eezh}, respectively.
Since $G_F$ is defined through the radiative correction to the muon decay,
which is a charged-current process, the $W$-boson propagator does not
receive radiative corrections in the orders of interest here.
Therefore, the correction factors of all these processes coincide with the one
for $\Gamma(H\to W^+W^-)$.

Finally, we shall treat the $ZZH$ interaction.
The procedure is very similar to the $W^+W^-H$ case.
Application of the low-energy theorem (\ref{let}) to the bare $Z$-boson vacuum
polarization induced by the top quark yields
\begin{equation}
{\cal L}_{ZZH}=(M_Z^0)^2Z_\mu^0Z^{\mu0}{H^0\over v^0}
\left[1-{(m_t^0)^2\partial\over\partial(m_t^0)^2}\,
{\Pi_{ZZ}(0)\over(M_Z^0)^2}\right].
\end{equation}
Again, $(M_Z^0)^2=M_Z^2+\delta M_Z^2$, with $\delta M_Z^2=\Pi_{ZZ}(0)$, and
$Z_\mu^0=Z_\mu$.
Together with Eq.~(\ref{du}), we then have
\begin{equation}
\label{lzzh}
{\cal L}_{ZZH}=2^{1/4}G_F^{1/2}M_Z^2Z_\mu Z^\mu H(1+\delta_{ZZH}),
\end{equation}
where $\delta_{ZZH}=\delta_u+\delta_{nu}^{ZZH}$,
with the non-universal part,
\begin{equation}
\delta_{nu}^{ZZH}=\left[1-{(m_t^0)^2\partial\over\partial(m_t^0)^2}\right]
{\Pi_{ZZ}(0)\over(M_Z^0)^2},
\end{equation}
being separately finite, gauge independent, and RG invariant.

Starting from the expression for $\Pi_{ZZ}(0)$ listed in Ref.~\cite{avd}
and repeating the steps of the $W^+W^-H$ analysis, we obtain
\begin{eqnarray}
\bar\delta_{ZZH}&\n=\n&N_cx_t\left\{-{5\over6}
+hC_F\left(-5l-2\zeta(2)+{25\over3}\right)
+h^2C_F\left[243S_2-{449\over6}\zeta(3)-{14\over3}\zeta(2)+{79\over3}
\right.\right.
\nonumber\\
&\n+\n&C_F\left(-15l^2+l\left(-12\zeta(2)+{155\over2}\right)+4B_4+2D_3
-{1053\over2}S_2+2\zeta(4)+{383\over3}\zeta(3)
\right.\nonumber\\
&\n-\n&\left.
{259\over9}\zeta(2)+{593\over72}\right)
+C_A\left(-{55\over6}l^2+l\left(-{22\over3}\zeta(2)+{65\over18}\right)
-2B_4-D_3+{1053\over4}S_2
\right.\nonumber\\
&\n+\n&\left.
15\zeta(4)-{293\over6}\zeta(3)-{73\over3}\zeta(2)
+{613\over24}\right)
\nonumber\\
&\n+\n&\left.\left.
n_f\left({5\over3}l^2+l\left({4\over3}\zeta(2)-{25\over9}\right)
-{8\over3}\zeta(3)+{14\over3}\zeta(2)-{35\over12}\right)\right]\right\}
\nonumber\\
&\n\approx\n&
-{5\over6}N_cx_t\left[1+a(2\,l-2.017\,386)
+a^2\left({15\over4}l^2-4.815\,198\,l-1.086\,685\right)\right]
\end{eqnarray}
in the $\overline{\mbox{MS}}$ scheme and
\begin{eqnarray}
\label{dzzhos}
\delta_{ZZH}&\n=\n&N_cX_t\left\{-{5\over6}
+hC_F\left(-2\zeta(2)+15\right)
+h^2C_F\left[243S_2-{449\over6}\zeta(3)+{46\over3}\zeta(2)+{49\over3}
\right.\right.
\nonumber\\
&\n+\n&C_F\left(4B_4+2D_3-{1053\over2}S_2+2\zeta(4)+{443\over3}\zeta(3)
+\zeta(2)\left(-80\ln2+{335\over9}\right)-{1019\over9}\right)
\nonumber\\
&\n+\n&C_A\left(L\left(-{22\over3}\zeta(2)+55\right)
-2B_4-D_3+{1053\over4}S_2+15\zeta(4)-{353\over6}\zeta(3)
\right.\nonumber\\
&\n+\n&\left.\left.\left.\!
\zeta(2)\left(40\ln2-{113\over3}\right)+{3697\over36}\right)
+n_f\left(L\left({4\over3}\zeta(2)-10\right)
-{8\over3}\zeta(3)-2\zeta(2)-{115\over9}\right)\right]\right\}
\nonumber\\
&\n\approx\n&-{5\over6}N_cX_t[1-4.684\,053\,a-a^2(8.197\,092\,L+6.846\,779)]
\end{eqnarray}
in the on-shell scheme.
The ${\cal O}(G_FM_t^2)$ and ${\cal O}(\alpha_sG_FM_t^2)$ terms of
Eq.~(\ref{dzzhos}) agree with those found in Refs.~\cite{hzz,ks2},
respectively.
Again, the ${\cal O}(\alpha_s^2G_FM_t^2)$ term of Eq.~(\ref{dzzhos}) reinforces
the potential of the QCD corrections to reduce the leading-order $M_t$
dependence.
Comparing $\delta_{ZZH}$ with $\Delta\rho$, $\delta_u$, and $\delta_{WWH}$,
we observe that it has the largest $\alpha_s/\pi$ coefficient but the smallest
$(\alpha_s/\pi)^2$ coefficient, the ratio of the latter to the former only
being about 1.5.
The $(\alpha_s/\pi)^2$ coefficient of $\delta_{ZZH}$ is by 53\% smaller than
the one of $\Delta\rho$ in Eq.~(\ref{drhoos}).

{}From Eq.~(\ref{lzzh}), we infer that $\Gamma(H\to ZZ)$ receives the
correction factor
\begin{equation}
\label{kzzh}
K_{ZZH}=1+2\delta_{ZZH}.
\end{equation}
The Born formula for $\Gamma(H\to ZZ)$ and its full one-loop
correction may be found in Ref.~\cite{hzz}.
Since the condition $2M_Z<M_H\ll M_t$ is likely to be unrealistic \cite{abe},
the high-$M_t$ approximation underlying Eq.~(\ref{lzzh}) is of limited
usefulness for $H\to ZZ$.
We can evade this problem by allowing for one or both of the $Z$ bosons to go
off-shell.
In addition to the information contained in Eq.~(\ref{lzzh}), we then need to
account for the corresponding corrections arising from the gauge sector.
However, in order not to invoke unknown QCD corrections, we have to restrict
ourselves to the inclusion of lepton lines.
The form of the additional corrections depends on the considered reaction.
It is useful to divide the phenomenologically relevant processes into three
classes:
\begin{itemize}
\item[(1)] $H\to Z^*Z\to f\bar fZ$, $Z\to Z^*H\to f\bar fH$,
and $e^+e^-\to ZH$;
\item[(2)] $H\to Z^*Z^*\to f\bar ff^\prime\bar f^\prime$ and
$e^+e^-\to Z^*\to Z^*H\to f\bar fH$ (via Higgs-strahlung);
\item[(3)] $e^+e^-\to e^+e^-Z^*Z^*\to e^+e^-H$ (via $ZZ$ fusion).
\end{itemize}
Here, $f$ and $f^\prime$ stand for neutrinos and charged leptons.
The results
for $H\to f\bar fZ$ at tree level \cite{riz} and at one loop \cite{pr},
for $Z\to f\bar fH$ at tree level \cite{zhgg} and at one loop \cite{zffh},
for $e^+e^-\to ZH$ at tree level and at one loop \cite{eezh},
for $H\to f\bar ff^\prime\bar f^\prime$ at tree level \cite{hzgg},
and for $e^+e^-\to f\bar fH$ at tree level \cite{eezh}
are in the literature.
In the next section, we shall discuss the corrections to these processes in
${\cal O}(\alpha_s^nG_FM_t^2)$, with $n=0,1,2$.

\section{Corrections from the gauge sector}

The IBA \cite{iba} provides a
systematic and convenient method to incorporate the dominant corrections of
fermionic origin to processes within the gauge sector of the SM.
These are contained in $\Delta\rho$ and
$\Delta\alpha=1-\alpha/\overline\alpha$, which parameterizes the running of the
fine-structure constant from its value, $\alpha$, defined in Thomson scattering
to its value, $\overline\alpha$, measured at the $Z$-boson scale.
The recipe is as follows.
Starting from the Born formula expressed in terms of $\alpha$, $c_w$, $s_w$,
and the physical particle masses, one substitutes
\begin{eqnarray}
\alpha&\n\to\n&\overline\alpha={\alpha\over1-\Delta\alpha},\nonumber\\
c_w^2&\n\to\n&\overline c_w^2=1-\overline s_w^2=c_w^2(1-\Delta\rho).
\end{eqnarray}
To eliminate $\overline\alpha$ in favour of $G_F$, one exploits the relation
\begin{equation}
{\sqrt2\over\pi}G_F={\overline\alpha\over\overline s_w^2M_W^2}
={\overline\alpha\over\overline c_w^2\overline s_w^2M_Z^2}(1-\Delta\rho),
\end{equation}
which correctly accounts for the leading fermionic corrections.

We shall now employ the IBA to find the additional corrections through
${\cal O}(\alpha_s^2G_FM_t^2)$ to the four- and five-point processes with a
$ZZH$ coupling, which we have classified in Section~2.
We shall always assume that the Born formulae are written in terms of
$G_F$, $c_w$, $s_w$, and the physical particle masses.
The generic correction factor for class~(1) reads \cite{ks2,zffh}
\begin{eqnarray}
\label{k1}
K_1^{(f)}&\n=\n&{(1+\delta^{ZZH})^2\over1-\Delta\rho}\,
{\overline v_f^2+a_f^2\over v_f^2+a_f^2}\nonumber\\
&\n=\n&1+2\delta^{ZZH}+
\left(1-8c_w^2{Q_fv_f\over v_f^2+a_f^2}\right)\Delta\rho,
\end{eqnarray}
where $v_f=2I_f-4s_w^2Q_f$, $\overline v_f=2I_f-4\overline s_w^2Q_f$,
$a_f=2I_f$, $Q_f$ is the electric charge of $f$ in units of the positron
charge, $I_f$ is the third component of weak isospin of the left-handed
component of $f$, and we have omitted terms of ${\cal O}(G_F^2M_t^4)$ in the
second line.
Similarly, the correction factor for class~(2) is given by \cite{gro}
\begin{eqnarray}
\label{k2}
K_2^{(ff^\prime)}&\n=\n&
{(1+\delta^{ZZH})^2\over(1-\Delta\rho)^2}\,
{\overline v_f^2+a_f^2\over v_f^2+a_f^2}\,
{\overline v_{f^\prime}^2+a_{f^\prime}^2\over v_{f^\prime}^2+a_{f^\prime}^2}
\nonumber\\
&\n=\n&1+2\delta^{ZZH}+
2\left[1-4c_w^2\left({Q_fv_f\over v_f^2+a_f^2}
+{Q_{f^\prime}v_{f^\prime}\over v_{f^\prime}^2+a_{f^\prime}^2}\right)\right]
\Delta\rho.
\end{eqnarray}
Here and in the following, we neglect interference terms of five-point
amplitudes with a single fermion trace, since, in the kinematic regime of
interest here, these are strongly suppressed, by $\Gamma_V/M_V$, with $V=W,Z$.
Such terms have recently been included in a tree-level calculation of
$\Gamma(H\to2V\to4f)$ for $M_H\ll M_W$ \cite{asa}.

The correction factor for case~(3) is slightly more complicated because the
electron and positron lines run from the initial state to the final state.
Allowing for generic fermion flavours, $f$ and $f^\prime$,
the Born cross section may be evaluated from
\begin{equation}
\label{zfus}
\sigma(ff^\prime\to ff^\prime H)={G_F^3M_Z^4\over64\pi^3\sqrt2}
\left[(v_f^2+a_f^2)(v_{f^\prime}^2+a_{f^\prime}^2)A
\pm4v_fa_fv_{f^\prime}a_{f^\prime}B\right],
\end{equation}
where
\begin{equation}
\label{int}
A=\int_{M_H^2/s}^1dx\int_x^1dy{a(x,y)\over\left[1+s(y-x)/M_Z^2\right]^2},
\end{equation}
and similarly for $B$, $\sqrt s$ is the centre-of-mass energy, and the
plus/minus sign refers to an odd/even number of antifermions in the initial
state.
The process under case~(3), with an $e^+e^-$ initial state, requires the plus
sign.
The integrands read
\begin{eqnarray}
a(x,y)&\n=\n&
\left({2x\over y^3}-{1+2x\over y^2}+{2+x\over2y}-{1\over2}\right)
\left[{z\over1+z}-\ln(1+z)\right]
+{x\over y^2}\left({1\over y}-1\right){z^2\over1+z},\nonumber\\
b(x,y)&\n=\n&
\left(-{x\over y^2}+{2+x\over2y}-{1\over2}\right)
\left[{z\over1+z}-\ln(1+z)\right],
\end{eqnarray}
where $z=(y/M_Z^2)(s-M_H^2/x)$.
The inner integration in Eq.~(\ref{int}) has been carried out analytically
in Appendix~A of Ref.~\cite{eezh}.
By means of the IBA, we obtain the correction factor for Eq.~(\ref{zfus}) as
\begin{eqnarray}
\label{k3}
K_3^{(ff^\prime)}&\n=\n&
{(1+\delta^{ZZH})^2\over(1-\Delta\rho)^2}\,
{(\overline v_f^2+a_f^2)(\overline v_{f^\prime}^2+a_{f^\prime}^2)A
\pm4\overline v_fa_f\overline v_{f^\prime}a_{f^\prime}B\over
(v_f^2+a_f^2)(v_{f^\prime}^2+a_{f^\prime}^2)A
\pm4v_fa_fv_{f^\prime}a_{f^\prime}B}\nonumber\\
&\n=\n&1+2\delta^{ZZH}+
2\left[1-{4c_w^2\over1+r}\left({Q_fv_f\over v_f^2+a_f^2}
+{Q_{f^\prime}v_{f^\prime}\over v_{f^\prime}^2+a_{f^\prime}^2}\right)
\right.\nonumber\\&\n-\n&\left.
{2c_w^2\over1+1/r}\left({Q_f\over v_f}+{Q_{f^\prime}\over v_{f^\prime}}\right)
\right]\Delta\rho,
\end{eqnarray}
where
\begin{equation}
r={\pm4v_fa_fv_{f^\prime}a_{f^\prime}B\over
(v_f^2+a_f^2)(v_{f^\prime}^2+a_{f^\prime}^2)A}.
\end{equation}
We wish to point out that, in the limit $r\to0$, $K_3$ coincides with $K_2$.
Detailed analysis reveals that $r$ is quite small in magnitude whenever
$e^+e^-\to e^+e^-H$ via $ZZ$ fusion is phenomenologically relevant.
In fact, if we consider energies $\sqrt s>150$~GeV and demand that the total
cross section of this process be in excess of $10^{-2}$~fb$^{-1}$, then we
find $|r|<1\%$.
This concludes our discussion of the additional QCD corrections to the
processes under items~(1)--(3) originating in the gauge sector.

\section{Numerical results}

We are now in a position to explore the phenomenological implications of our
results.
We shall take the values of our input parameters to be
$M_W=80.26$~GeV,
$M_Z=91.1887$~GeV \cite{lep},
$M_t=180$~GeV \cite{abe},
and $\alpha_s^{(5)}(M_Z)=0.118$ \cite{bet}.\footnote{Note that this value does
not include results from lattice computations.}
The latter corresponds to $\alpha_s^{(6)}(M_t)=0.1071$, which entails that
$\Lambda_{\overline{MS}}^{(6)}=91$~MeV in Eq.~(\ref{as}).
If we use the one-loop formula for $\alpha_s^{(6)}(\mu)$, {\it i.e.},
Eq.~(\ref{as}) with the second term within the square brackets discarded,
$\Lambda_{\overline{MS}}^{(6)}$ comes down to 41~MeV.

Any perturbative calculation to finite order depends on the choice of
renormalization scheme and, in general, also on one or more renormalization
scales.
It is generally believed that the scheme and scale dependences of a
calculation up to a given order indicate the size of the unknown higher-order
contributions, {\it i.e.}, they provide us with an estimate of the theoretical
uncertainty.
Of course, the central values and variations of the scales must be judiciously
chosen in order for this estimate to be meaningful.
If the perturbation series converges, then the scheme and scale dependences
are expected to decrease as the respective next order is taken into account.
This principle has recently been confirmed for $\Delta\rho$ \cite{avd,yr}.
Here, we have the opportunity to carry out similar studies for the three
additional observables $\delta_u$, $\delta_{WWH}$, and $\delta_{ZZH}$.
Similarly to Ref.~\cite{avd}, we have presented our results in the on-shell
and $\overline{\mbox{MS}}$ schemes as functions of a single renormalization
scale, $\mu$.
In the $\overline{\mbox{MS}}$ scheme, one could, in principle, introduce
individual renormalization scales for the coupling and the mass.
{}For simplicity, we have chosen not to do so.
It is natural to define the central value of $\mu$ in such a way that, at
this point, the radiative correction is devoid of logarithmic terms.
This leads us to set $\mu=\xi M_t$ in the on-shell scheme and $\mu=\xi\mu_t$,
where $\mu_t=m_t(\mu_t)$, in the $\overline{\mbox{MS}}$ scheme.
We may obtain $\mu_t$ as a closed function of $M_t$ by iterating
Eq.~(\ref{mass}), with the result that
\begin{eqnarray}
\label{mut}
{\mu_t\over M_t}&\n=\n&1-4HC_F
+H^2C_F\left\{-12\zeta(2)+6
+C_F\left[-12\zeta(3)+6\zeta(2)(8\ln2-5)+{199\over8}\right]
\right.\nonumber\\
&\n+\n&\left.
C_A\left[6\zeta(3)+8\zeta(2)(-3\ln2+1)-{1111\over24}\right]
+n_f\left[4\zeta(2)+{71\over12}\right]\right\}
\nonumber\\
&\n\approx\n&1-{4\over3}A-6.458\,784\,A^2,
\end{eqnarray}
where $A=4H=\alpha_s(M_t)/\pi$.
{}For $M_t=180$~GeV, Eq.~(\ref{mut}) yields $\mu_t=170.5$~GeV, in good
agreement
with the exact fix point of Eq.~(\ref{mass}), which is $\mu_t=170.6$~GeV.

\begin{table}[ht]
\caption{Relative deviations (in \%) of $\Delta\rho$, $\delta_u$,
$\delta_{WWH}$, and $\delta_{ZZH}$ from the respective one-loop results due to
their corrections up to ${\cal O}(\alpha_s)$ and ${\cal O}(\alpha_s^2)$.
The renormalization scale dependence is investigated by choosing $\mu=\xi M_t$,
with $\xi$ variable.
}\label{tab:os}
\medskip
\begin{tabular}{|c|c|c|c|c|c|c|c|c|} \hline\hline
\rule{0mm}{5mm}$\xi$ &
\multicolumn{2}{c|}{$\Delta\rho/\Delta\rho^{(1)}-1$ [\%]} &
\multicolumn{2}{c|}{$\delta_u/\delta_u^{(1)}-1$ [\%]} &
\multicolumn{2}{c|}{$\delta_{WWH}/\delta_{WWH}^{(1)}-1$ [\%]} &
\multicolumn{2}{c|}{$\delta_{ZZH}/\delta_{ZZH}^{(1)}-1$ [\%]} \\ \cline{2-9}
\rule{0mm}{5mm} &
${\cal O}(\alpha_s)$ & ${\cal O}(\alpha_s^2)$ &
${\cal O}(\alpha_s)$ & ${\cal O}(\alpha_s^2)$ &
${\cal O}(\alpha_s)$ & ${\cal O}(\alpha_s^2)$ &
${\cal O}(\alpha_s)$ & ${\cal O}(\alpha_s^2)$ \\ \hline
1/4 & $-11.68$ & $-11.88$ & $ -7.34$ & $ -8.65$ & $ -9.33$ & $ -9.35$ &
$-19.13$ & $-16.58$ \\
1/2 & $-10.63$ & $-11.72$ & $ -6.68$ & $ -8.34$ & $ -8.49$ & $ -9.24$ &
$-17.40$ & $-16.83$ \\
1 & $ -9.75$ & $-11.44$ & $ -6.12$ & $ -8.01$ & $ -7.78$ & $ -9.04$ &
$-15.96$ & $-16.76$ \\
2 & $ -9.00$ & $-11.11$ & $ -5.66$ & $ -7.67$ & $ -7.19$ & $ -8.79$ &
$-14.74$ & $-16.51$ \\
4 & $ -8.36$ & $-10.74$ & $ -5.26$ & $ -7.35$ & $ -6.68$ & $ -8.51$ &
$-13.70$ & $-16.15$ \\
\hline\hline
\end{tabular}
\end{table}

\begin{table}[ht]
\caption{Relative deviations (in \%) of $\Delta\bar\rho$, $\bar\delta_u$,
$\bar\delta_{WWH}$, and $\bar\delta_{ZZH}$ from the respective one-loop results
due to their corrections up to ${\cal O}(\alpha_s)$ and ${\cal O}(\alpha_s^2)$.
The renormalization scale dependence is investigated by choosing
$\mu=\xi\mu_t$, with $\xi$ variable.
}\label{tab:ms}
\medskip
\begin{tabular}{|c|c|c|c|c|c|c|c|c|} \hline\hline
\rule{0mm}{5mm}$\xi$ &
\multicolumn{2}{c|}{$\Delta\bar\rho/\Delta\rho^{(1)}-1$ [\%]} &
\multicolumn{2}{c|}{$\bar\delta_u/\delta_u^{(1)}-1$ [\%]} &
\multicolumn{2}{c|}{$\bar\delta_{WWH}/\delta_{WWH}^{(1)}-1$ [\%]} &
\multicolumn{2}{c|}{$\bar\delta_{ZZH}/\delta_{ZZH}^{(1)}-1$ [\%]} \\
\cline{2-9}
\rule{0mm}{5mm} &
${\cal O}(\alpha_s)$ & ${\cal O}(\alpha_s^2)$ &
${\cal O}(\alpha_s)$ & ${\cal O}(\alpha_s^2)$ &
${\cal O}(\alpha_s)$ & ${\cal O}(\alpha_s^2)$ &
${\cal O}(\alpha_s)$ & ${\cal O}(\alpha_s^2)$ \\ \hline
1/4 & $-15.63$ & $-11.15$ & $-10.70$ & $ -8.25$ & $-12.96$ & $ -8.67$ &
$-24.09$ & $-15.15$ \\
1/2 & $-10.89$ & $-11.55$ & $ -6.87$ & $ -8.22$ & $ -8.71$ & $ -9.09$ &
$-17.77$ & $-16.56$ \\
1 & $ -8.96$ & $-11.19$ & $ -5.63$ & $ -7.79$ & $ -7.15$ & $ -8.86$ &
$-14.69$ & $-16.51$ \\
2 & $ -8.64$ & $-10.88$ & $ -5.82$ & $ -7.56$ & $ -7.11$ & $ -8.71$ &
$-13.49$ & $-16.16$ \\
4 & $ -9.24$ & $-10.85$ & $ -6.81$ & $ -7.70$ & $ -7.92$ & $ -8.84$ &
$-13.41$ & $-15.93$ \\
\hline\hline
\end{tabular}
\end{table}

In Tables~\ref{tab:os} and \ref{tab:ms}, we investigate the $\xi$ dependence of
$\delta_u$, $\delta_{WWH}$, and $\delta_{ZZH}$ and their $\overline{\mbox{MS}}$
counterparts, respectively.
{}For comparison, we also include the results for $\Delta\rho$ and
$\Delta\bar\rho$.
To be specific, we consistently evaluate these quantities to leading and
next-to-leading order in QCD and study their relative deviations from their
respective one-loop values, which we denote by the superscript (1),
{\it e.g.}, $\Delta\rho^{(1)}=N_cX_t$, {\it etc.}
Notice that the on-shell and $\overline{\mbox{MS}}$ results coincide at one
loop.
In our ${\cal O}(\alpha_s)$ analysis, we use the one-loop formula for
$\alpha_s(\mu)$ with $\Lambda_{\overline{MS}}^{(6)}=41$~MeV and omit the
${\cal O}(\alpha_s^2)$ terms in Eqs.~(\ref{mass}) and (\ref{mut}).
{}For the time being, let us concentrate on the entries for $\xi=1$ and
assess the effect of the QCD corrections as well as their scheme dependence.
We observe that, in both schemes, the QCD corrections are throughout negative,
even for $\bar\delta_u$ and $\bar\delta_{WWH}$, where the
${\cal O}(\alpha_s)$ and ${\cal O}(\alpha_s^2)$ terms are in part positive.
This is due to the fact that we consistently compute all QCD parameters,
{\it i.e.}, $\alpha_s(\mu)$, $m_t(\mu)$, and $\mu_t$, to the orders under
consideration.
The reduction in $x_t$, which occurs as an overall factor in the
$\overline{\mbox{MS}}$ formulae, happens to overcompensate the positive
effect of these particular coefficients.
Inclusion of the ${\cal O}(\alpha_s^2)$ terms in
$(\Delta\rho,\delta_u,\delta_{WWH},\delta_{ZZH})$ increases the size of the
QCD corrections by $(17,31,16,5)\%$, respectively.
In the $\overline{\mbox{MS}}$ case, the increments amount to
$(25,38,24,12)\%$ of the respective ${\cal O}(\alpha_s)$ corrections.
As might be expected, the scheme dependence of the QCD corrections to this
quadruplet of quantities is dramatically reduced, by $(68,55,71,80)\%$, as we
pass from ${\cal O}(\alpha_s)$ to ${\cal O}(\alpha_s^2)$.
Let us now also include the other $\xi$ values in our consideration.
Within each scheme, we determine the scale dependence of the QCD correction to
a given quantity by comparing its largest and smallest values in the interval
$1/4\le\xi\le4$.
As expected, the scale dependence is drastically decreased when we take the
${\cal O}(\alpha_s^2)$ terms into account, namely by $(66,37,68,87)\%$ and
$(89,87,93,86)\%$ in the on-shell and $\overline{\mbox{MS}}$ schemes,
respectively.
The exceptionally small reduction of the scale dependence in the case of
$\delta_u$ is due to the fact that $\delta_u$ has the smallest
${\cal O}(\alpha_s)$ term and the largest ${\cal O}(\alpha_s^2)$ term of all
four on-shell quantities.

\begin{table}[ht]
\caption{Coefficients of the correction factors in the form of Eq.~(\ref{kfac})
for the various Higgs-boson decay rates and production cross sections discussed
in the text.
In the last line, $x=B/A$, where $A$ and $B$ are given by Eq.~(\ref{int}),
and terms of ${\cal O}(x^2)$ have been neglected.
}\label{tab:k}
\medskip
\begin{tabular}{|c|c|c|c|} \hline\hline
$K$ & $C_1$ & $C_2$ & $C_3$ \\ \hline
$K_{\ell\ell H}$ & $7/3$ & $-1.797$ & $-16.201$ \\
$K_{WWH}$ & $-5/3$ & $-2.284$ & $-10.816$ \\
$K_{ZZH}$ & $-5/3$ & $-4.684$ & $-6.847$ \\
$K_1^{(\nu)}$ & $-2/3$ & $-7.420$ & $4.774$ \\
$K_1^{(\ell)}$ & $-1.272$ & $-5.249$ & $-4.445$ \\
$K_2^{(\nu\nu)}$ & $1/3$ & $6.261$ & $-53.330$ \\
$K_2^{(\nu\ell)}$ & $-0.272$ & $-14.025$ & $32.824$ \\
$K_2^{(\ell\ell)}$ & $-0.878$ & $-6.323$ & $0.113$ \\
$K_3^{(\ell\ell)}$ & $-0.878-2.353\,x$ & $-6.323+9.281\,x$ &
 $0.113-39.416\,x$ \\
\hline\hline
\end{tabular}
\end{table}

In the remainder of this section, we shall stick to the on-shell scheme.
In Eqs.~(\ref{kllh}), (\ref{kwwh}), (\ref{kzzh}), (\ref{k1}), (\ref{k2}), and
(\ref{k3}), we have presented correction factors for various Higgs-boson
production cross sections and decay rates in terms of $\Delta\rho$, $\delta_u$,
$\delta_{WWH}$, and $\delta_{ZZH}$.
It is instructive to cast these correction factors into the generic form
\begin{equation}
\label{kfac}
K=1+C_1\Delta\rho^{(1)}\left[1+C_2a\left(1+{7\over4}aL\right)+C_3a^2\right],
\end{equation}
where $C_i$ $(i=1,2,3)$ are numerical coefficients.
Notice that we have kept the full $\mu$ dependence in Eq.~(\ref{kfac}).
We could have written Eqs.~(\ref{drhoos}), (\ref{duos}), (\ref{dwwhos}), and
(\ref{dzzhos}) in the same way.
The fact that the coefficient of $aL$ is universal may be understood by
observing that $K$ represents a physical observable, which must be RG invariant
through the order of our calculation, and that, to leading order of QCD, $K$
only implicitly depends on $\mu$, via $a$.
In fact, the coefficient of $aL$ is nothing but $\beta_0$ of Eq.~(\ref{beta}).
The outcome of this decomposition is displayed in Table~\ref{tab:k}.
In the case of $K_3^{(\ell\ell)}$, we have treated $x=B/A$, where $A$ and $B$
are defined in Eq.~(\ref{int}), as an additional expansion parameter and
discarded terms of ${\cal O}(x^2)$.
This is justified because, in practice, $|x|\ll1$,
{\it e.g.}, for $\sqrt s=300$~GeV and $M_H=100$~GeV, we find
$x\approx-5.233\cdot10^{-2}$.
While in the case of the three basic corrections, $K_{\ell\ell H}$,
$K_{WWH}$, and $K_{ZZH}$, $C_2$ and $C_3$ are both negative, this is not
in general so.
In fact, in all composite corrections, except for $K_1^{(\ell)}$, the
${\cal O}(\alpha_s^2)$ terms partially compensate the ${\cal O}(\alpha_s)$
ones.
In $K_2^{(\nu\nu)}$, we even find a counterexample to the heuristic rule
\cite{ks2} that, in the $G_F$ formulation of the on-shell scheme, the
${\cal O}(G_FM_t^2)$ terms get screened by their QCD corrections.
In the latter case, we also encounter a gigantic value of $C_3$.
Both features may be ascribed to the fact, that, in ${\cal O}(G_FM_t^2)$, the
$\delta_{ZZH}$ and $\Delta\rho$ terms of $K_2^{(\nu\nu)}$ largely cancel.
The extraordinarily large value of $C_3$ in $K_2^{(\nu\ell)}$ is also
accompanied by a suppression of $C_1$.
The $C_1$ values of $K_2^{(\ell\ell)}$ and $K_3^{(\ell\ell)}$ are relatively
small, too.
We are thus in the fortunate position that the leading high-$M_t$ corrections
to the $2\to3$ and $1\to4$ processes of Higgs-boson production and decay with a
$ZZH$ coupling, for which full one-loop calculations have not yet been
performed, are throughout quite small.
Thus, there is hope that the subleading fermionic corrections to these
processes will not drastically impair the situation.
However, the IBA does not provide us with any information on the bosonic
corrections.

\begin{table}[ht]
\caption{Full one-loop weak corrections (in \%) to various Higgs-boson decay
rates and production cross sections and their ${\cal O}(G_FM_t^2)$ terms.
In the last line, we have used $\protect\sqrt s=175$~GeV.
}\label{tab:full}
\medskip
\begin{tabular}{|c|c|c|c|} \hline\hline
Observable & $M_H$ [GeV] & ${\cal O}(\alpha)$ weak [\%] &
 ${\cal O}(G_FM_t^2)$ [\%] \\ \hline
$\Gamma(H\to\tau^+\tau^-)$ & 75 & 1.792 & 2.369 \\
$\Gamma(H\to\nu\bar\nu Z)$ & 105 & 1.275 & $-0.677$ \\
$\Gamma(H\to\ell^+\ell^-Z)$ & 105 & $-1.220$ & $-1.292$ \\
$\Gamma(Z\to\nu\bar\nu H)$ & 65 & 0.024 & $-0.677$ \\
$\Gamma(Z\to\ell^+\ell^-H)$ & 65 & 0.296 & $-1.292$ \\
$\sigma(e^+e^-\to ZH)$ & 75 & $-2.293$ & $-1.292$ \\
\hline\hline
\end{tabular}
\end{table}

In this context, it is interesting to revisit processes for which the full
one-loop weak corrections are known and to investigate in how far the
${\cal O}(G_FM_t^2)$ terms play a dominant r\^ole there.
Here, we are only interested in reactions which already proceed at tree level.
Specifically, we shall consider
$Z\to f\bar fH$ \cite{zffh} for $M_H=65$~GeV,
$H\to\tau^+\tau^-$ \cite{hff} and $e^+e^-\to ZH$ \cite{eezh} for $M_H=75$~GeV,
and $H\to f\bar fZ$ \cite{pr} for $M_H=105$~GeV, where $f=\nu,\ell$.
Our analysis of $\sigma(e^+e^-\to ZH)$ will refer to LEP2 energy,
$\sqrt s=175$~GeV.
In all these cases, the quantumelectrodynamical (QED) and weak corrections are
separately finite and gauge independent at one loop.
In Table~\ref{tab:full}, we compare the full one-loop weak corrections to these
processes with their ${\cal O}(G_FM_t^2)$ terms.
In the case of $H\to\tau^+\tau^-$, $H\to\ell^+\ell^-Z$, and $e^+e^-\to ZH$,
the ${\cal O}(G_FM_t^2)$ terms give a reasonably good account of the full
corrections, while they come out with the wrong sign in the other cases.
However, the full calculations for $M_H=65$~GeV give very small results
anyway.
On the other hand, the ${\cal O}(G_FM_t^2)$ term for $H\to\nu\bar\nu Z$ is
suppressed due to a partial cancellation between $\delta_{ZZH}$ and
$\Delta\rho$ in $K_1^{(\nu)}$ and cannot be expected to dominate the full
correction.
Whenever the full correction is known, it should be included on the right-hand
side of Eq.~(\ref{kfac}) with the ${\cal O}(G_FM_t^2)$ term subtracted.
In conclusion, the radiative corrections considered in Table~\ref{tab:full} all
appear to be well under control.

\section{Conclusions}

In this paper, we have presented the three-loop ${\cal O}(\alpha_s^2G_FM_t^2)$
corrections to the effective Lagrangians for the interactions of light Higgs
bosons with pairs of charged leptons, $W$ bosons, and $Z$ bosons in the SM.
While the demand for corrections in this order is certainly more urgent in the
gauge sector \cite{avd}, where precision test are presently being carried out,
our analysis is also interesting from a theoretical point of view, since it
allows us to recognize a universal pattern.
In addition to $\Delta\rho$, we have now three more independent observables
with quadratic $M_t$ dependence at our disposal for which the QCD expansion is
known up to next-to-leading order, namely $\delta_u$, $\delta_{WWH}$, and
$\delta_{ZZH}$.
In the on-shell scheme of electroweak and QCD renormalization, these four
electroweak parameters exhibit striking common properties.
In fact, the leading- and next-to-leading-order QCD corrections act in the
same direction and screen the ${\cal O}(G_FM_t^2)$ terms.
Even the sets of $\alpha_s/\pi$ and $(\alpha_s/\pi)^2$ coefficients each
lie in the same ball park.
{}For the choice $\mu=M_t$, the coefficients of $\alpha_s/\pi$ range between
$-1.797$ and $-4.684$, and those of $(\alpha_s/\pi)^2$ between $-6.847$ and
$-16.201$.
If we compare this with the corresponding coefficients of the ratio
$\mu_t^2/M_t^2$, which are $-2.667$ and $-11.140$,
then it becomes apparent that the use of the top-quark pole mass is the origin
of these similarities.
Here, $\mu_t=m_t(\mu_t)$, for which we have presented a closed two-loop
formula.
If we express the QCD expansions in terms of $\mu_t$ rather than $M_t$ and
choose $\mu=\mu_t$,
then the coefficients of $\alpha_s/\pi$ and $(\alpha_s/\pi)^2$ nicely group
themselves around zero;
they range from $-2.017$ to 0.870 and from $-3.970$ to $1.344$, respectively.
This indicates that the perturbation expansions converge more rapidly if we
renormalize the top-quark mass according to the $\overline{\mbox{MS}}$ scheme.
Without going into details, we would like to mention that the study of
renormalons \cite{ren} offers a possible theoretical explanation of this
observation.
Since the on-shell and $\overline{\mbox{MS}}$ results coincide in lowest
order, this does, of course, not imply that the QCD corrections are any
smaller in the $\overline{\mbox{MS}}$ scheme.
It just means that, as a rule, the ${\cal O}(G_FM_t^2)$ terms with $M_t$
replaced by the two-loop expression for $\mu_t$ are likely to provide fair
approximations for the full three-loop results.
Furthermore, we have demonstrated that, similarly to $\Delta\rho$, the scheme
and scale dependences of $\delta_u$, $\delta_{WWH}$, and $\delta_{ZZH}$ are
considerably reduced when the next-to-leading-order QCD corrections are taken
into account.
Armed with this information, we have made rather precise predictions for a
variety of production and decay processes of low-mass Higgs bosons at present
and future $e^+e^-$ colliders.
In all the cases considered here, the radiative corrections appear to be well
under control now.

\bigskip
\centerline{\bf ACKNOWLEDGEMENTS}
\smallskip\noindent

We would like to thank Bill Bardeen, Kostja Chetyrkin, and Michael Spira for
very useful discussions.
One of us (BAK) is indebted to the FNAL Theory Group for inviting him as a
Guest Scientist.
He is also grateful to the Phenomenology Department of the University of
Wisconsin at Madison for the great hospitality extended to him during a
visit when a major part of his work on this project was carried out.

\newpage

\end{document}